%% file: main.tex
\documentclass[letterpaper, 10 pt, conference]{ieeeconf}   

\IEEEoverridecommandlockouts                              

\usepackage{graphics} 
\usepackage{epsfig} 
\usepackage{amsmath} 
\usepackage{amssymb}  
\usepackage{amsthm}
\usepackage{mathtools}
\usepackage{color}
\usepackage{tikz}
\usepackage{tikzsymbols}
\usetikzlibrary{shapes,arrows,backgrounds,arrows.meta,positioning,calc}
\tikzstyle{block} = [draw, rectangle,
  minimum height=1cm, minimum width=3cm, text width=2.5cm, align=center]
\theoremstyle{plain}
\newtheorem{assumption}{Assumption}
\newtheorem{definition}{Definition}

\newcommand{\mUpred}{u}

\newcommand{\mXpred}{x}

\usepackage[implicit=false]{hyperref}

\title{\LARGE \bf
A predictive safety filter for learning-based racing control
}

\author{Ben Tearle and Kim P. Wabersich and Andrea Carron and Melanie N. Zeilinger
\thanks{$^{1}$The authors are with the Institute for Dynamical Systems and Control, ETH Zurich, ZH-8092, Switzerland:
        {\tt\small bentearle@gmail.com, [wkim|carrona|mzeilinger]@ethz.ch}. 
        This work was supported by the Swiss National Science Foundation under grant no. PP00P2 157601 / 1. The research of Andrea Carron was supported by the Swiss National Centre of Competence in Research NCCR Digital Fabrication.
        The work of Ben Tearle and Andrea Carron was supported by the ETH Career Seed Grant 19-18-2.} %
}

\begin{document}

\maketitle
\thispagestyle{empty}
\pagestyle{empty}

\begin{abstract}
	The growing need for high-performance controllers in safety-critical applications like autonomous driving has been motivating the development of formal safety verification techniques.
	In this paper, we design and implement a predictive safety filter that is able to maintain vehicle safety with respect to track boundaries when paired alongside any potentially unsafe control signal, such as those found in learning-based methods.
	A model predictive control (MPC) framework is used to create a minimally invasive algorithm that certifies whether a desired control input is safe and can be applied to the vehicle, or that provides an alternate input to keep the vehicle in bounds.
	To this end, we provide a principled procedure to compute a safe and invariant set for nonlinear dynamic bicycle models using efficient convex approximation techniques.
	To fully support an aggressive racing performance without conservative safety interventions, the safe set is extended in real-time through predictive control backup trajectories.
	Applications for assisted manual driving and deep imitation learning on a miniature remote-controlled vehicle demonstrate the safety filter's ability to ensure vehicle safety during aggressive maneuvers.
\end{abstract}

\section{INTRODUCTION}

The development of robotic systems has led to an ever increasing number of applications that go beyond the isolated task spaces found in legacy industries such as automotive or electronics production.
More recent applications encompass dynamic and learning-based interactions with humans in complex task spaces, as is the case with autonomous driving, and therefore require advanced safety mechanisms~\cite{Ames2019,Hewing2020}, to prevent potentially dangerous situations.
Maintaining safety at the physical limits for highly dynamic systems often requires a task-specific trade-off between performance and conservatism to ensure safe system operation.
As a result, there is an increasing interest in developing theoretically sound safety frameworks with a reduced degree of conservatism that enable safety in a modular fashion, independent of a task-specific objective.

While some of these methods have been demonstrated in practice, the considered applications are often small-scale or nearly linear control systems that are only operated within conservative regions of their state space~\cite{Fisac2019}.
Motivated by the strict safety requirements in autonomous driving, we consider the problem of safe autonomous and assisted racing as a benchmark application for deriving a practically relevant safety mechanism.
Racing requires the utilization of a vehicle's full nonlinear dynamics, providing a challenging domain in which safety must be guaranteed.
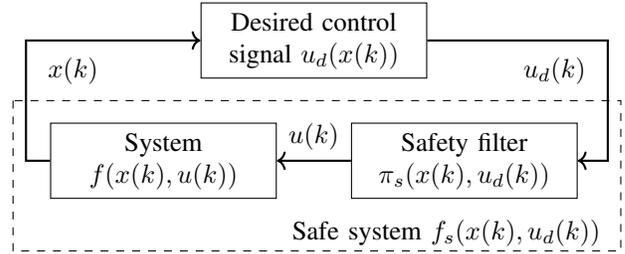
\begin{figure}
	\centering
	\begin{tikzpicture}[scale=0.8]
		\input{fig/tikz/controlLoopSafe.tex}
	\end{tikzpicture}
	\caption{Concept of predictive safety filter: Based on the current state
		$x(k)$, an arbitrary control algorithm provides a desired control input
		$u_d(k)\in\mathbb R^m$, which is processed by the safety filter
		$u(k) = \pi_s(x(k),u_d(k))$ and applied to the real system.}
	\label{fig:concept}
\end{figure}

To provide safety for arbitrary control policies, we rely on a modular safety framework as shown in Figure~\ref{fig:concept}.
This approach allows the framework to be used in conjunction with any potentially unsafe control signal, such as those from learning-based controllers.
The basic idea is to design a \emph{safety filter}, which analyzes the desired control signal and decides in real-time whether it can be applied to the system, or if it has to be modified to ensure safety.
For the racing application considered in this work, this consists of verifying if the vehicle is able to stay within track boundaries in the future given the current steering and drivetrain commands.

This is achieved by finding safe backup control sequences that lead the vehicle towards a set of known safe states, where the first input of the sequence is as close as possible to the desired control signal.
This approach allows for verifying the safety of the desired input while at the same time providing an alternative safe input otherwise.

\subsection{Related Work}
The concept of using a safety controller in a closed-loop system was first introduced in~\cite{Seto1998}, where the system can switch between an experimental controller and a reliable safety controller in the case of software faults.
Developments on the theoretic use of barrier certificates for verifying system safety were later proposed in~\cite{Prajna2004}, which was further extended to the idea of control barrier functions (CBFs)~\cite{Wieland2007}.
More recent work has revisited the notion of using CBFs for safety-critical control of robotic systems, see~\cite{Ames2019} for an overview.
This approach has been combined with a machine-learning framework in~\cite{taylor2019learning} to safely learn model discrepancies of a Segway robot while limiting the operational space during training.
Although these methods build off strong theoretical results from control Lyapunov function theory, they rely on the ability to explicitly model a system's safety requirements as a CBF, which is not generally trivial to design.

Given the inherent lack of safety guarantees in traditional machine learning methods, the reinforcement learning (RL) field has become increasingly interested in enforcing constraints for training black-box control policies.
A general-purpose policy search algorithm for constrained reinforcement learning is introduced in~\cite{achiam2017constrained}, which approximately enforces safety constraints at every policy update. Using a learning-based system model,
\cite{Berkenkamp2016a} proposes a method for determination of a safe set of system states under a specific learning-based policy.
Although these methods allow for approximately safe policy training, they are limited in that they remain tied to task-specific reinforcement learning algorithms, whereas the safety filter presented in this work is able to function independent of a specific task and thereby enables modular safety.

An approach for providing system safety based on confining a system to a pre-computed set of safe states is introduced in~\cite{Gillula2011}.
This uses reachability-based techniques to find a safe set for a given system together with a corresponding control policy that provides invariance within the safe set.
The idea is expanded in~\cite{Fisac2019} to perform online updates of the safe set using a non-parametric system dynamics estimate.
These approaches suffer from limited scalability in the offline safe set computation required.
Recent work attempts to address this by approximating the reachable sets using data-based methods~\cite{wabersich2017scalableSafety}, sum-of-squares programming~\cite{Wang2018a}, and active learning~\cite{Chakrabarty2020}.

Closely related to these ideas, a method for establishing safety using an MPC-based control law is derived in~\cite{Wabersich2019}.
A continuously updating control policy is computed online to find backup trajectories towards safe states, resulting in the implicit representation of the safe set and corresponding safe control law via the MPC optimization problem.
This method is extended to consider nonlinear stochastic systems formulated with chance-constraints or parametric uncertainties in~\cite{Wabersich2021probabilistic,Wabersich2021}, and provides the foundation for the task of autonomous racing considered in this work.

\subsection{Contributions}
The main contribution of this paper is the design and implementation of a permissive safety filter for autonomous racing that can be combined with any desired control signal, ensuring closed-loop vehicle safety with respect to a track for a diverse range of applications.
To this end, we use the concept of predictive safety filters as presented in \cite{Wabersich2019,Wabersich2021}.
To achieve a minimally invasive safety filter supporting aggressive maneuvers, we use a nonlinear dynamic bicycle model with a Pacejka model of the tire forces~\cite{pacejka2002tyre} to simultaneously predict and optimize accurate backup control trajectories.
In addition to a high-fidelity system model, the safety filter performance can be improved by using either a longer planning horizon or a larger terminal set.
As the planning horizon is typically limited by memory and processing requirements,
we derive an iterative optimization-based invariant set computation using convex approximations to obtain an enlarged terminal safe set for the nonlinear dynamic bicycle model, which is valid over a range of constant road curvatures.

The physical miniature racing application demonstrates the proposed safety filter's performance with both human-in-the-loop racing and deep imitation learning.
This work presents, to the best of our knowledge, the first application of a predictive safety filter to a complex and highly dynamical nonlinear system demonstrated in experimental results.

\section{PROBLEM FORMULATION}

\emph{Notation}: The set of integers in the interval $[a, b] \subset \mathcal{R}$ is denoted by $\mathcal{I}_{[a, b]}$, and the set of integers in the interval $[a,\infty) \subset \mathcal{R}$ is $\mathcal{I}_{\geq a}$.
The $i$-th row of a matrix $M \in \mathcal{R}^{n \times m}$ is denoted by $[M]_i$.

The goal of this work is to design a safety filter that certifies whether or not a desired control input, $u_d(k)$, is safe for a vehicle system, and provides an alternative safe control input at any time.
We consider a discrete-time nonlinear system of the form
\begin{equation}
	\label{eq:general_system}
	x(k+1) = f(x(k), u(k)), ~ \forall k\in \mathcal{I}_{\geq 0},
\end{equation}
subject to state and input constraints, $x(k)\!\in\!\mathcal{X}, u(k)\!\in\!\mathcal{U}$, where the dynamics $f:\mathcal X \times \mathcal U \rightarrow \mathbb R^n$.
System safety is defined with respect to ensuring constraint satisfaction at all times, as follows.
\begin{definition}\label{def:safe}
	A system~\eqref{eq:general_system} is considered safe if
	\begin{equation}\label{eq:safe}
		x(k) \in \mathcal{X}, u(k) \in \mathcal{U}, \ \forall k \in \mathcal{I}_{\geq 0}.
	\end{equation}
\end{definition}

In order to guarantee this notion of safety for a given $u_d(k)$, a safety control policy, $\pi_{\mathcal{S}}(x(k), u_d(k))$, is provided that guarantees constraint satisfaction for all future timesteps if applied to the vehicle.
If a safety policy exists with $u_d(k)$ as the current input of the policy, then $u_d(k)$ can be certified as safe and applied to the system.
More formally:

\begin{definition}\label{def:input_safety}
	A desired input $u_d(\bar{k})$ is \it{certified as safe} for system~\eqref{eq:general_system}, at a given timestep $\bar{k}$, if the safety control policy yields $\pi_{\mathcal{S}}(x(\bar{k}), u_d(\bar{k}))=u_d(\bar{k})$, and application of $u(k)=\pi_{\mathcal{S}}(x(k), u_d(k))$ to the system results in safety according to Definition~\ref{def:safe} for all $k \geq \bar{k}$.
\end{definition}
Using a safety policy in accordance with Definition~\ref{def:input_safety} provides a safety filter that can be brought into a closed-loop system as shown in Figure~\ref{fig:concept}.
Since the safety policy can be updated at each time step to consider the incoming desired input, this allows the desired control signal to have control authority over the system whenever possible, i.e. $\pi_{\mathcal{S}}(x(k), u_d(k))\!=\!u_d(k)$.
However, if the desired control signal would put the system at risk of violating its constraints in the future, then alternate inputs, $\pi_{\mathcal{S}}(x(k), u_d(k))\! \neq \!u_d(k)$, must be available that ensure safety for the system.

The next section discusses an approach to compute~$\pi_{\mathcal{S}}$ online using an MPC framework that minimizes interference while still ensuring safety for the system.

\section{PREDICTIVE SAFETY FILTER}\label{sec:safety_filter}

\begin{figure*}[t]
	\centering{
		\vspace{5pt}
		\includegraphics[width=\textwidth]{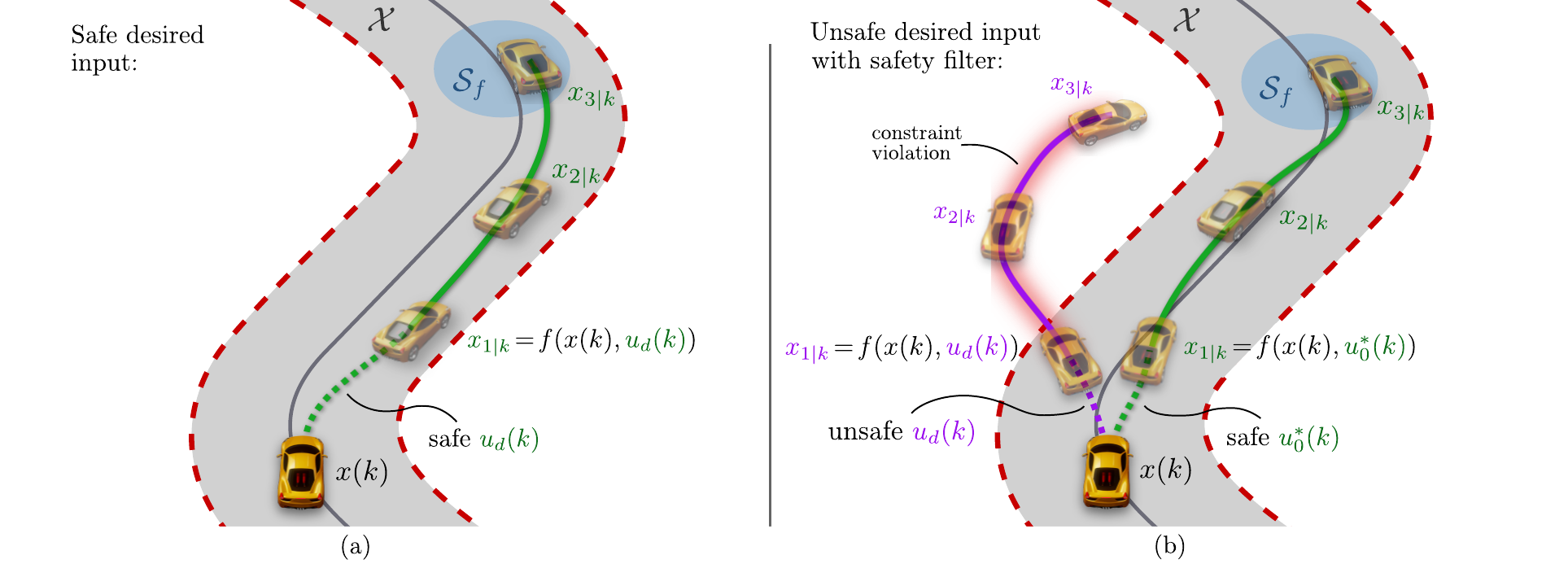}
		\caption{Diagram (a) shows a possible vehicle trajectory from a safe desired input $u_d(k)$. Diagram (b) shows the resulting vehicle trajectory from an unsafe desired input, where the vehicle ends up leaving the track. An alternate safe input $u^*_0(k)$ applied by the safety filter is shown along with its trajectory.}
		\label{fig:safety_diagram}
	}
\end{figure*}

We define an implicit safety policy through a receding-horizon optimal
control problem, referred to as \emph{predictive safety filter problem}~\cite{Wabersich2019},
which allows for an efficient online computation of the desired safety filter $\pi_\mathcal{S}$:

\begin{subequations}\label{eq:MPSF_nominal_opt}
	\begin{align} \label{eq:MPSF_nominal_opt_cost}
		\min_{\substack{\mXpred_{i|k},\mUpred_{i|k}}} ~ & J(u_{i|k}, u_d(k))                                                                                  \\
		\nonumber \text{s.t.} ~ \quad                   & \forall i \in \mathcal{I}_{[0, N-1]}:                                                               \\
		                                                & \mXpred_{0|k} = x(k), \label{eq:MPSF_nominal_opt_initial_constraint}                                \\
		                                                & \mXpred_{i+1|k} = f(\mXpred_{i|k}, \mUpred_{i|k}),~ \label{eq:MPSF_nominal_opt_dynamic_constraints} \\
		                                                & \mXpred_{i|k} \in \mathcal X,~ \label{eq:MPSF_nominal_opt_state_constraints}                        \\
		                                                & \mUpred_{i|k} \in \mathcal U,~ \label{eq:MPSF_nominal_opt_input_constraints}                        \\
		                                                & \mXpred_{N|k} \in \mathcal S_f.  \label{eq:MPSF_nominal_opt_safety_constraint}
	\end{align}
\end{subequations}
Problem~\eqref{eq:MPSF_nominal_opt} computes a discrete-time state and input backup trajectory, $\{ x_{i|k}^*, u_{i|k}^* \}$, of length $N$, where $x_{i|k}$ is the state predicted $i$ timesteps ahead, computed at time $k$, initialized at $x_{0|k} = x(k)$, and similarly for $u_{i|k}$.
The system is predicted along the horizon according to dynamics~\eqref{eq:MPSF_nominal_opt_dynamic_constraints}, subject to an initial condition~\eqref{eq:MPSF_nominal_opt_initial_constraint}, state and input constraints~\eqref{eq:MPSF_nominal_opt_state_constraints} and~\eqref{eq:MPSF_nominal_opt_input_constraints}, and terminal constraint~\eqref{eq:MPSF_nominal_opt_safety_constraint}.
Different from classical MPC, the objective function in~\eqref{eq:MPSF_nominal_opt_cost} is chosen to minimize the difference between the desired control input and the first input of the solution trajectory, as
\begin{equation}\label{eq:cost_fn}
	J(u_{i|k}, u_d(k)) = \Vert u_d(k) - \mUpred_{0|k}\Vert^2.
\end{equation}
The safety policy is then defined by $\pi_{\mathcal{S}}(x(k), u_d(k)) = u_{0|k}^*$.

The cost function in~\eqref{eq:cost_fn} can be modified to include secondary objectives beyond tracking the desired control signal.
For the racing application, we include a regularization term that penalizes the rate of change of the inputs in order to encourage a smoother control trajectory:
\begin{align}
	J(u_{i|k}, u_d(k)) = \Vert u_d(k) - \mUpred_{0|k}\Vert^2_W  + \sum_{i=0}^{N-1} \Vert \Delta u_{i\vert k} \Vert^2_{R_{\mathcal{S}}},
\end{align}
where $\Delta u_{0|k}:=u_{0|k}-u_{0|k-1}$, $\Delta u_{i|k}:=u_{i|k}-u_{i-1|k}$ for $i=1,..,N-1$, and $W, R_{\mathcal{S}} \in \mathbb{R}^{m \times m}$ are cost matrices for the input deviation and input rate respectively.
This helps to reduce rapid fluctuations between the desired input and safety filter's input, which can occur with the system at the boundary of the state constraints in practice.
To avoid unnecessary input deviations from a desired input that can be certified as safe, the weights are chosen
with $W$ much larger than $R_{\mathcal{S}}$ to ensure priority remains on tracking the desired input.

\begin{assumption}[Invariant terminal set]\label{ass:terminal_invariant_set}
	There exists a control law $\kappa_f: \mathcal{S}_f \rightarrow {\mathcal U}$, and a corresponding positively invariant set $\mathcal{S}_f \subseteq \mathcal X$, such that for all $x \in \mathcal{S}_f$, it holds that $\kappa_f(x) \in \mathcal U$ and $f( x,\kappa_f(x) ) \in \mathcal{S}_f$.
\end{assumption}

As in standard MPC theory, Assumption~\ref{ass:terminal_invariant_set} provides recursive feasibility for the safety control policy obtained from problem~\eqref{eq:MPSF_nominal_opt}, i.e. if the problem has a feasible solution at timestep $\bar{k}$, then a feasible solution also exists for all future times $k > \bar{k}$.
This results in constraint satisfaction at all times, meeting the requirements for a safe system put forth in Definition~\ref{def:safe}.

If we consider the case where $u_d(k)$ is a safe input for the system, there must exist a state and input trajectory, beginning at $x(k\!+\!1)\!=\!f(x(k), u_d(k))$, that is feasible along the horizon and ends in $\mathcal{S}_f$.
An example is shown in Figure~\ref{fig:safety_diagram}(a), where a vehicle is at initial state $x(k)$, and the primary state constraint is to stay inside track limits.
Application of the input $u_d(k)$ would bring the vehicle to state $x_{1|k}$, from where a state and input trajectory exists that keeps the vehicle inside the boundaries before reaching $\mathcal{S}_f$.
The input $u_d(k)$ can therefore be certified as safe, and the optimal solution to~\eqref{eq:MPSF_nominal_opt} would be $u_{0|k}^*\!=\!u_d(k)$.
This achieves a minimal objective cost of zero, satisfying the desired behavior of no intervention for a safe $u_d(k)$.

If the desired input is unsafe, then any resulting trajectory beginning at $x(k\!+\!1)\!=\!f(x(k), u_d(k))$ must violate the constraints at some point along the horizon.
Looking at Figure~\ref{fig:safety_diagram}(b), the trajectory following $x_{1|k}$ after applying $u_d(k)$ can be seen to leave the track.
In this case, Problem~\eqref{eq:MPSF_nominal_opt} will provide an input, $u_{0|k}^*\! \neq \! u_d(k)$, that is able to maintain system safety while being as close as possible to $u_d(k)$.
A backup control trajectory is shown in the same figure that can be taken instead if a safe initial input is applied.

\section{VEHICLE DYNAMICS AND CONSTRAINTS}
In this section, the model used to describe the vehicle dynamics is presented, followed by the system constraints.

\subsection{System Model}
\begin{figure}[h]
	\centering{
		\includegraphics[scale=1.0]{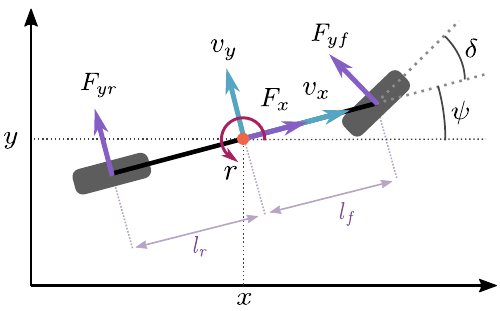}
		\caption{Dynamic vehicle model diagram.}
		\label{fig:dynamic_model}
	}
\end{figure}

In this work we consider a miniature RC car, which is modeled using a standard dynamic bicycle model formulation \cite{rajamani2011vehicle,Liniger14}.
Using a dynamic model as opposed to a simpler kinematic model as considered in previous related work, see, e.g., \cite{Ames2019}, allows us to consider the nonlinear tire forces which have a significant impact on vehicle motion during aggressive maneuvers.
The state of the model is
$x = \left[p_x, p_y, \psi, v_x, v_y, r \right],$
with the input
$u = \left[\delta, \tau \right],$
where $p_x$, $p_y$ are the x-y coordinates of the car and $\psi$ is the heading angle in the global coordinate frame; $v_x$, $v_y$, and $r$ are the velocities and yaw rate of change in the vehicle's body frame.
Finally, $\delta$ is the steering angle and $\tau$ is the drivetrain command.
An illustration can be seen in Figure~\ref{fig:dynamic_model}.

The system model can be described by the differential equations
\begin{equation}
	\label{eq:model}
	\dot{x} =
	\begin{bmatrix}
		v_x \cos(\psi) - v_y \sin(\psi)                                \\
		v_x \sin(\psi) - v_y \cos(\psi)                                \\
		r                                                              \\
		\frac{1}{m} \left(F_x - F_{yf}\sin(\delta) + m v_y r\right)    \\
		\frac{1}{m} \left(F_{yr} + F_{yf}\cos(\delta) - m v_x r\right) \\
		\frac{1}{I_z} \left(F_{yf}l_f\cos(\delta) - F_{yr}l_r \right)
	\end{bmatrix},
\end{equation}
where $m$ is the car mass, $I_z$ is the yaw moment of inertia, and $l_{f/r}$ is the distance between the center of gravity and the front and rear axles, respectively.
The lateral tire forces $F_{yf}$ and $F_{yr}$ are modeled with a simplified Pacejka tire model,
\begin{equation}
	\label{eq:lateral_forces}
	\begin{split}
		&\alpha_f = \arctan\left(\frac{v_y + l_f r}{v_x}\right) - \delta,~~\alpha_r = \arctan\left(\frac{v_y - l_r r}{v_x}\right)\\
		&F_{yf/yr} = D_{f/r} \sin(C_{f/r}\arctan(B_{f/r}\alpha_{f/r})),
	\end{split}
\end{equation}
where $\alpha_f$ and $\alpha_r$ are the tire slip angles \cite{pacejka2002tyre}.
The longitudinal force is modeled as a single force applied to the center of gravity of the vehicle, and is computed as a linear combination of the drivetrain command and velocity as  $F_x = C_1\tau + C_2\tau^2 + C_3 v_x + C_4 v_x^2 + C_5 \tau v_x$.
The drivetrain command $\tau$ can be positive, resulting in forward motion, or negative, resulting in braking.

The continuous-time system in~\eqref{eq:model} is discretized using Euler forward, obtaining a discrete-time nonlinear system of the form~\eqref{eq:general_system}.

\subsection{System Constraints}

The system is subject to nonlinear state constraints, and polyhedral input constraints of the form
\begin{equation}
	\label{eq:constraints}
	\mathcal X := \{x \in \mathbb R^n \vert d(x) \leq b \},\quad
	\mathcal U := \{u \in \mathbb R^m \vert G u \leq g \},
\end{equation}
where $d : \mathbb{R}^{n} \rightarrow \mathbb{R}^{n_b}$, and $G \in \mathbb{R}^{n_g \times m}$.
The input constraints consist of bounding the maximum and minimum commands, while the state constraints enforce the safety-critical task of keeping the car within track limits.

\begin{figure}[h]
	\centering{
		\includegraphics[scale=1.9]{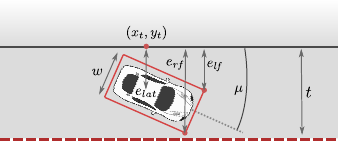}
		\caption{Track-relative error states used to constraint the vehicle.}
		\label{fig:e_lat}
	}
\end{figure}

To keep the vehicle within the boundaries of the track, we constrain the front two corners of a bounding box around the vehicle, $e_{lf}$ and $e_{rf}$, shown in Figure~\ref{fig:e_lat}.
The lateral error of the vehicle's center of gravity with respect to the track center-line is $e_{lat}$, while the yaw error of the vehicle with respect to the track orientation is $\mu$.
Given a reference center-line position and orientation, $x_{t}$, $y_{t},\psi_{t}$, these states can be written as
\begin{align}\label{eq:e_lat}
	\begin{split}
		e_{lat}(k) &= -\sin(\psi_{t})(x(k) - x_{t}) + \cos(\psi_{t})(y(k) - y_{t}),\\
		\mu(k) &= \psi(k) - \psi_{t},\\
		e_{lf}(k) &= e_{lat}(k) + l_f \sin(\mu(k)) + \frac{w}{2} \cos(\mu(k)), \\
		e_{rf}(k) &= e_{lat}(k) + l_f \sin(\mu(k)) - \frac{w}{2} \cos(\mu(k)),
	\end{split}
\end{align}
where $w$ is the width of the vehicle.
These two corner points of the bounding box can be bounded by half the width of the track, denoted $t$, as
\begin{equation}\label{eq:track_constraint}
	\left|e_{lf}\right| \leq t, \quad \left|e_{rf}\right| \leq t.
\end{equation}
\vspace{-0.7cm}

\section{TERMINAL SET COMPUTATION}
The main difficulty in designing a safety filter for the considered racing application is the construction of the positively invariant set, $\mathcal{S}_f$, for the nonlinear vehicle system as described in Assumption~\ref{ass:terminal_invariant_set}.
A method for computing polyhedral terminal sets for autonomous driving is presented in~\cite{LimaLTVMPC}, but the required simplifying assumptions in the kinematic model used are not suitable for a vehicle performing aggressive maneuvers.
Approaches to terminal set design for more general nonlinear systems can be found in~\cite{Chen,Carron20,CONTE2016117}, where the common idea is to design a set based on a linearized system while using techniques to compensate for linearization errors such that set invariance still holds for the nonlinear system.
We take a similar approach that enforces a required Lyapunov dissipation for a range of steady-states to compute the terminal safe set.

We first introduce a transformation in a track-relative coordinate frame that allows computing steady-states of the nonlinear vehicle model parameterized by the road curvature.
Based on established techniques for terminal set design, we then propose to compute a linear control law capable of stabilizing the nonlinear system in a neighborhood around a specific steady state.
We consider a grid of parameter values for the linearized system and compute a positively invariant set for track segments of constant curvature.
A-posteriori verification is then performed to ensure invariance holds for the nonlinear system across the full parameter range.

\subsection{Track-Relative Coordinate Transformation and Terminal Steady-States}
\label{subsec:track_relative_transformation}
For the safety certification problem presented in Section~\ref{sec:safety_filter}, the terminal set must contain states that are considered safe for the desired system.
In a racing context, having the vehicle positioned on the center-line and oriented forwards is a safe position, providing the vehicle is able to follow the center-line closely under some control law.
In order to more easily analyze the system with respect to the center-line, the global state is transformed into the track-relative state $x_{r} = \left[e_{lat}, \mu, v_{x}, v_{y}, r \right]$, similar to that used in \cite{vazquez2020optimizationbased}. Here, $e_{lat}$ and $\mu$ are the lateral error and orientation error as described in~\eqref{eq:e_lat}, and $v_{x}, v_{y}$, and $r$ remain unchanged from~\eqref{eq:model}.
The dynamics of $e_{lat}$ and $\mu$ are described by
\begin{align}\label{eq:track_relative}
	\begin{split}
		\dot{e}_{lat} &= v_x \sin(\mu) + v_y\cos(\mu), \\
		\dot{\mu} &= r - c\displaystyle\frac{v_x \cos(\mu) - v_y\sin(\mu)}{1-ce_{lat}},
	\end{split}
\end{align}
which are parameterized by the curvature of the track, $c$, at a given point on the center-line.
We use the same dynamics for $v_x,v_y,r$ as in~\eqref{eq:model} to describe $\dot{x}_r$, then discretize to obtain
\begin{equation}
	\label{eq:discrete_relative_system}
	x_{r}(k\!+\!1, c) = f_{r}\left(x_{r}(k,c), u(k)\right), ~ \forall k\in \mathcal{I}_{\geq 0},
\end{equation}
with $f_{r}:\mathbb{R}^{n_{r}} \times \mathbb{R}^{m} \rightarrow \mathbb R^{n_{r}}$.
Constraints keeping the vehicle within track boundaries, $|e_{lat}| \leq t-w/2$, and oriented forwards, $|\mu| \leq \pi/2$, can now be written in polytopic form as $\mathcal{X}_{r}\!\coloneqq\!\{x_{r}\!\in\!\mathbb R^{n_{r}} \vert H x_{r}\!\leq\!h \}$, where $H\!\in\!\mathbb{R}^{n_h \times n_r}$.

The goal is to find a terminal control law for the system~\eqref{eq:discrete_relative_system} that can stabilize the vehicle around the track center-line, relating to $e_{lat} = 0$ and a constant velocity $v_x = \textrm{v}_x$.
Since the track-relative dynamics are parameterized by $c$, different steady-state points $\left(x^{e}_{r}(c), u^{e}(c)\right)$ exist depending on the current track curvature.
The steady-state and corresponding input at a given curvature can be computed by solving~\eqref{eq:discrete_relative_system} for a state and input pairing such that $x_r^e(c)\!=\!f(x_r^e(c), u^e(c))$, resulting in
\begin{equation}
	x_r^e(c) =  [0, \mu^e, \textrm{v}_x, v_y^e, r^e]^T, \quad u^e(c) = [\delta^e, \tau^e]^T \label{eq:steady_state}.
\end{equation}
While direct use of the steady-state~\eqref{eq:steady_state} as a terminal constraint satisfies the invariance property, the
resulting terminal constraints~\eqref{eq:MPSF_nominal_opt_safety_constraint} would become rather restrictive, resulting
in conservative behavior of the safety filter. To increase the feasible set of~\eqref{eq:MPSF_nominal_opt} and thereby the safe set of the vehicle states, we  propose a design procedure to enlarge the terminal steady
state constraint through an invariant set in the following.

\subsection{Terminal Set \& Control Law Synthesis}
\label{subsec:terminal_set_controller}

To design a terminal set for the system~\eqref{eq:model}, we use a linearization around the previously introduced equilibrium points \eqref{eq:steady_state} to obtain a stabilizing state feedback controller.
This allows us to derive a positively invariant set from a Lyapunov function for the corresponding closed-loop system.

We begin by linearizing~\eqref{eq:discrete_relative_system} for a specific steady-state and curvature~\eqref{eq:steady_state}, resulting in
\begin{equation}\label{eq:linear_relative_system}
	\bar{x}_{r}(k+1, c) = A(c)\bar{x}_{r}(k, c) + B(c)\bar{u}(k, c)
\end{equation}
where $A(c)$ and $B(c)$ are the linearization matrices evaluated at a steady state pair $(x_r^e(c), u^e(c))$.
The notation $\bar{x}_r(k,c)=x_r(k,c)-x^e_r(c)$ indicates the deviation of the state $x_r(k,c)$ from the steady-state $x^e_r(c)$ for a given curvature, and similarly for $\bar{u}(k,c)$.
For the local stabilizing control law, we choose a constant linear controller of the form
\begin{equation}\label{eq:terminal_control_law}
	\kappa_f(k,c) = K\bar{x}_{r}(k, c),
\end{equation}
where $K \in \mathbb{R}^{m \times n_{r}}$.

An ellipsoidal set is chosen for the terminal set as
\begin{equation}\label{eq:terminal_set_ss}
	\mathcal{S}_f(c) \coloneqq \left\lbrace \bar{x}_{r}(k,c) \vert \bar{x}_r(k,c)^T P \bar{x}_r(k,c) \leq 1 \right\rbrace \subseteq \mathcal{X}_r,
\end{equation}
which is a sublevel set of a quadratic Lyapunov function $V_f(\bar{x}_r(k,c))\!=\!\bar{x}_r(k,c)^T P \bar{x}_r(k,c)$, contained within the state constraints $\mathcal{X}_r$.
The matrix $P\!\in\!\mathbb{R}^{n_r \times n_r}$ can be obtained by solving the discrete-time Lyapunov equation for the closed-loop system dynamics matrix $A_{cl}(c)\!=\!A(c)+B(c)K$, with a pre-specified dissipation rate $Q_{dis}$:
\begin{align}\label{eq:lyap_eq}
	A_{cl}(c)^T P A_{cl}(c)-P\leq -Q_{dis}.
\end{align}
The set~\eqref{eq:terminal_set_ss} is then guaranteed to be positively invariant for the system~\eqref{eq:linear_relative_system} at a given curvature when subject to the control law~\eqref{eq:terminal_control_law}.
The dissipation $Q_{dis}$ provides the ability to compensate for linearization errors when stabilizing the original nonlinear system.
This dissipation value is chosen using $Q_{dis}\!=\!Q+K^TRK$, where $Q, R$ are cost matrices that can be designed to bound the linearization errors by $\bar{x}_r(k,c)^T Q \bar{x}_r(k,c) + \bar{u}(k,c)^T R \bar{u}(k,c)$.

The curvature values of a track with both left and right turns fall into the range $c\!\in\![-c_{max}, c_{max}]$, where $c_{max}$ is the largest curvature value on the track.
We therefore want a single control law that stabilizes the system at any curvature within the given range.
This is done by first introducing a set of $n_{c} \in \mathbb{R}$ equidistant incremental curvature values in $[-c_{max}, c_{max}]$, and computing the corresponding equilibrium states, inputs, and linearization matrices for each: $\left\lbrace x_{r,i}^e, u_i^e, A_i, B_i \right\rbrace, \; \forall i \in \mathcal{I}_{[1,n_c]}$.
We then impose the stability condition from~\eqref{eq:lyap_eq} at each steady-state for the same control matrix $K$, computing the control law and resulting invariant set with a semidefinite program (similarly used in \cite{CONTE2016117}):
\begin{subequations}\label{eq:LMI_opt}
	\begin{align}
		\min_{E, Y}    & \quad -\log \det E \label{eq:LMI_opt_obj}                                                                 \\
		\textrm{s.t.}~ & ~ \forall i \in \mathcal{I}_{[1,n_c]} : \nonumber                                                         \\
		               & ~E \succeq 0                                                                                              \\
		               & \begin{bmatrix}
			\left([h]_j - [H]_j x^e_{r, i}\right)^2 & [H]_j E \\
			E [H]_j^T                               & E
		\end{bmatrix} \succeq 0, \forall j \in \mathcal{I}_{[1,n_h]}\label{eq:LMI_opt_x_constraints} \\
		               & \begin{bmatrix}
			\left([g]_l - [G]_l u^e_{i}\right)^2 & [G]_l E \\
			E [G]_l^T                            & E
		\end{bmatrix} \succeq 0, \forall l \in \mathcal{I}_{[1,n_g]}\label{eq:LMI_opt_u_constraints} \\
		               & \begin{bmatrix}
			E                & \star & \star & \star \\
			A_i E + B_i Y    & E     & 0     & 0     \\
			Q^{\frac{1}{2}}E & 0     & I     & 0     \\
			R^{\frac{1}{2}}Y & 0     & 0     & I
		\end{bmatrix} \succeq 0  \label{eq:invariance}
	\end{align}
\end{subequations}
where $ E \coloneqq P^{-1}$, and $Y \coloneqq KE$.
The solution to~\eqref{eq:LMI_opt} allows us to extract a maximal volume ellipsoidal set~\eqref{eq:terminal_set_ss} that is invariant for the closed-loop system $A_{cl}(c)$ at each of the $n_c$ gridded curvature values.
The matrix inequalities described by~\eqref{eq:LMI_opt_x_constraints} and~\eqref{eq:LMI_opt_u_constraints} impose the state and input constraints for each equilibrium point.
The constraint in~\eqref{eq:invariance} can be derived from the Lyapunov decrease condition~\eqref{eq:lyap_eq} and Schur complements; the matrix is symmetric with $\star$ representing the corresponding transposed terms.

Since the resulting set is invariant for only the linearized system at the chosen curvature values by design, we must further verify that invariance holds for the nonlinear system across the continuous range of curvatures.
This is done via an additional optimization problem that searches the set for any state and curvature pairing that leads to an invariance violation for the nonlinear system under the computed terminal control law:
\begin{subequations}\label{eq:verification_opt}
	\begin{align}
		\max_{\bar{x}_r, c} \quad & \bar{x}_r(k+1, c)^{T} P \bar{x}_r(k+1, c) \label{eq:verification_opt_obj} \\
		\quad\textrm{s.t.} \quad  & \bar{x}_r(k, c)^{T} P \bar{x}_r(k, c) \leq 1                              \\
		                          & \bar{x}_r(k+1, c) = f(\bar{x}_r(k), \kappa_f(k), c)                       \\
		                          & c \in [c^{min}, c^{max}].
	\end{align}
\end{subequations}
If the optimal objective value~\eqref{eq:verification_opt_obj} is less than 1, then $\mathcal{S}_f(c)$ is verified as invariant for the nonlinear system; otherwise, the problem has found a state for which the set is not invariant under the nonlinear dynamics. In this case, the set can be incrementally scaled down until no violating points are found, with the limit reaching the vehicle steady-state as a feasible solution.

Note that the invariance guarantees of the proposed terminal set are valid for constant curvatures.
Since we consider a track made up of connecting constant curvature segments, the theoretical invariance property therefore holds on each individual segment.
However, the guarantees do not strictly hold for the instantaneous change of curvature between segments due to the resulting shift in steady-state set point.
Since the linearization-based control law and invariant set design inherently introduce some conservatism, we observe in practice that changing set points can still be efficiently compensated.
We therefore do not explicitly account for this change in curvature, and consider invariance for the individual track segments as practically adequate for the terminal safe set.

The curvature value used for the terminal set when solving Problem~\eqref{eq:MPSF_nominal_opt} is taken as the curvature a certain distance ahead of the vehicle along the track.
This distance is heuristically chosen as a function of the current desired torque input, $u_{d,\tau}(k)$, and the time horizon of the problem, $t_N \!=\!N \cdot T_{s}$, to generate a reasonable distance ahead for the terminal set.

\section{EXPERIMENTS}
To demonstrate the performance of the proposed safety filter, the scheme is implemented with a small remote controlled vehicle on a track, where the vehicle must stay inside track boundaries.
We first present an experiment showing the safety filter in a driver-assistance scenario, where the desired inputs are provided directly by a human driver.
This is followed by an example of a learning-based control application using imitation learning, where a neural network policy is safely learned and deployed on the vehicle.
A video of the experiments performed can be found at:\\
\centerline{\url{https://youtu.be/Aaly_IwQmfc}.}

\subsection{Problem Implementation}
To ensure feasibility of the MPC problem~\eqref{eq:MPSF_nominal_opt}, the track width constraint~\eqref{eq:track_constraint} and the terminal set constraint~\eqref{eq:terminal_set_ss}, are implemented as soft constraints.
The problem is solved online using acados \cite{verschueren2019acados} with a real-time iteration SQP scheme, horizon length of $N\!=\!60$, and sampling frequency of 80 Hz.
The terminal set computation in~\eqref{eq:LMI_opt} is solved offline using MOSEK~\cite{mosek}, with $n_c = 21$ equilibrium points spanning curvatures $[-2.5, 2.5]$.
The verification problem in~\eqref{eq:verification_opt} is solved 1000 times from randomly selected initial conditions, and the resulting objective value never exceeds~1.

\subsection{Experimental Platform}
A Kyosho Mini-Z 1:28-scale remote controlled vehicle is used on a 0.80 m constant-width track as the test platform for all experiments.
A VICON motion capture system provides vehicle position and orientation information, which is used by an Extended Kalman Filter to produce a complete state estimate.
The safe control inputs are sent via radio controller to the vehicle.
The closed loop system is implemented using ROS (Robotic Operating System) running on a Lenovo ThinkPad P1 with Ubuntu 18.04, Intel Core i7-9750H processor, and 32 GB RAM.

\subsection{Manual Driver Assistance}
By combining the safety certification with human driver inputs, a driver-assistance system is created that provides necessary intervention should the driver make a mistake that would endanger the vehicle.
Since the safety certification is designed to be minimally invasive, it gives the driver free control of the vehicle as long as their actions remain safe, only intervening when required.

\begin{figure}[h]
	\centering{
		\includegraphics[scale=1]{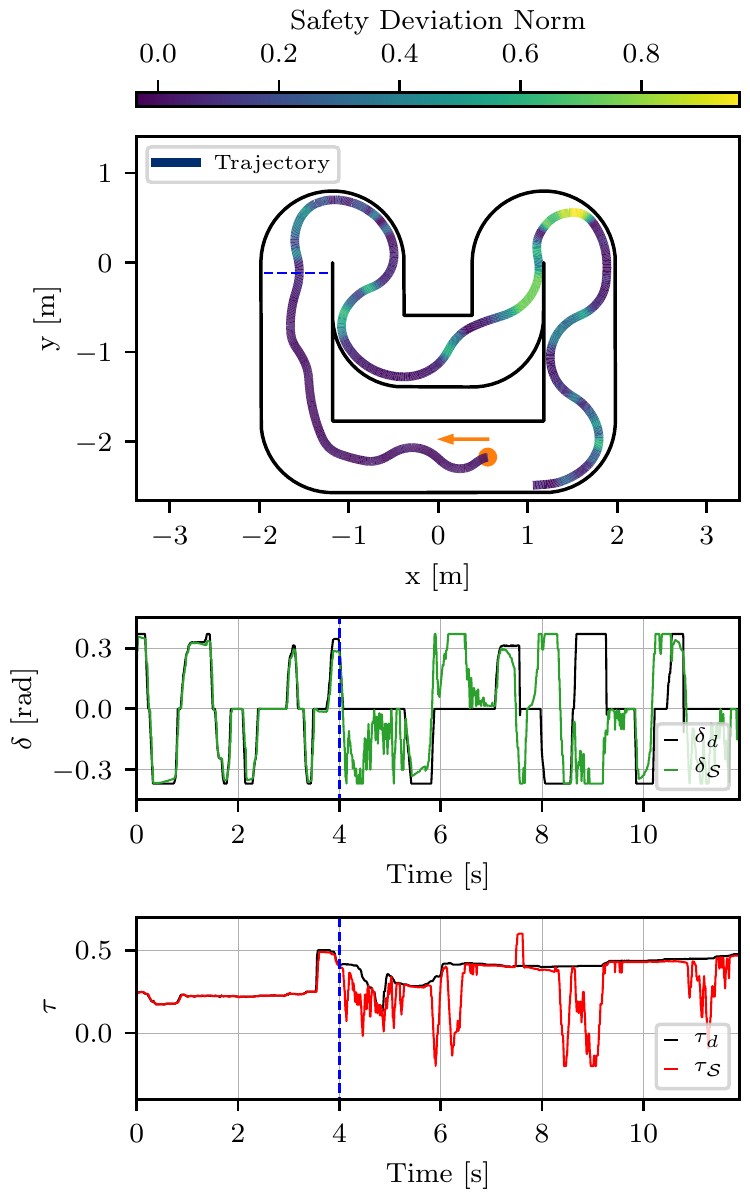}
		\caption{Vehicle trajectory (top) and control inputs (middle, bottom) for a human providing the desired control signal by joystick.
			The safety filter intervention is shown via heat map on the trajectory.
			The orange dot and arrow indicate the starting point and travel direction; the dashed blue line indicates the transition from generally safe driver inputs to unsafe inputs.}
		\label{fig:joystick_demo}
	}
\end{figure}

In this experiment, the manual driver inputs are provided by a physical joystick.
Figure~\ref{fig:joystick_demo} shows the vehicle trajectory and corresponding inputs from a single lap driven with the safety certification active.
In the vehicle trajectory plot, the color-map shows the L2-norm of the difference between the desired and safe control input vectors, indicating the magnitude of modification by the safety filter.
The input comparison plots show the safety filter commands initially closely correspond to the driver commands up until the dashed line, indicating that the driver commands are being certified as safe and applied to the vehicle.
After this, the safety certification begins to intervene in both steering and throttle inputs as the driver purposefully fails to steer around corners, or swerves the car toward the wall.
The plot of the trajectory demonstrates how the safety certification is able to keep the vehicle within track boundaries at all times, while still managing to track the desired inputs whenever possible.

\subsection{Imitation Learning}
Imitation learning is a technique for learning a policy that replicates the actions of a demonstration from another agent, typically an expert policy for the given task.
We implement an iterative algorithm called DAgger (Dataset Aggregation) \cite{RossDAgger}, to learn a stationary deterministic policy using imitation learning.
In DAgger, a policy is first initialized using supervised learning from an expert demonstration, and is then deployed directly on the task.
The expert labels all states visited by the learned policy with the optimal action, which is then added to the dataset for the policy to retrain on.
This process is repeated iteratively with the intention that the learned policy is able to improve from previous mistakes.

Since DAgger relies on rolling out the learner policy during training, it can be combined with the proposed safety filter to provide a safe training environment to learn a racing controller using the physical vehicle.
A feedforward neural network with 3 hidden layers, 64 neurons per layer, and ReLU activation functions is used as the policy architecture, which outputs a drivetrain and steering command. The input to the network is chosen as the vehicle state in track-relative coordinates, along with 30 curvature values over the next 1.5 meters of track as~$x_{NN}\!=\! \left[e_{lat},\mu,v_x,v_y,r,c_{1} \dots c_{30} \right]$.
Training the network consists of supervised learning to minimize the L2-norm of the difference between expert and network commands.
The expert policy used is a Model Predictive Contouring Controller (MPCC), presented in \cite{Liniger14}, which maximizes track progress while staying inside track boundaries, and has proved successful in other racing applications \cite{Kabzan19}.
Imitating a finite-horizon optimal policy like MPCC can be beneficial, as the states visited by the network controller each iteration can be labeled offline using an MPCC with a longer horizon that cannot be used in practice due to solve time requirements.
The resulting neural network then imitates a high performance policy that otherwise could not be achieved by the expert in real-time.

\begin{figure}[h]
	\centering{
		\includegraphics[scale=1]{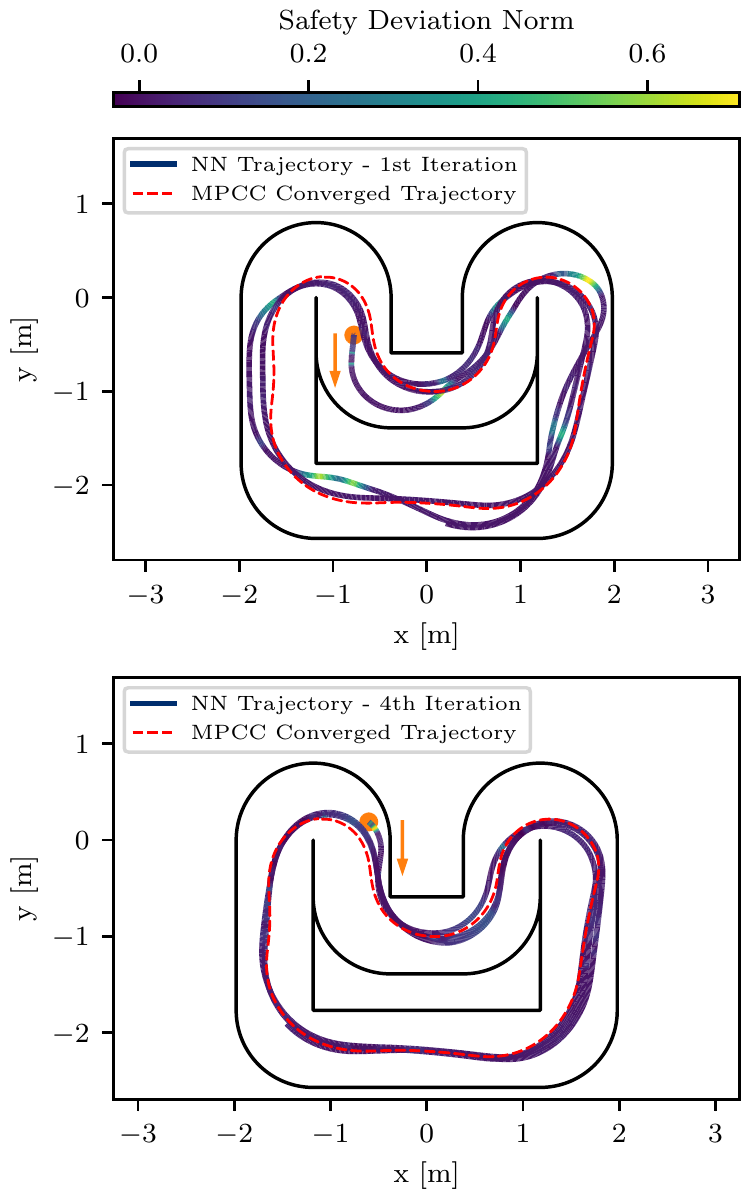}
		\caption{Vehicle trajectories shown during the first (top) and fourth (bottom) episodes of DAgger; safety filter intervention is shown via heat-map, and the converged expert MPCC trajectory is shown in red. Initial location and direction of travel are shown in orange.}
		\label{fig:dagger_demo}
	}
\end{figure}

DAgger is set up on the experimental platform alongside the safety filter to allow for completely automated safe training.
Safety is provided both during data collection when the neural network policy is operating, and during transition periods as the vehicle stops to retrain the policy.
Figure~\ref{fig:dagger_demo} shows two trajectory plots of DAgger episodes while the neural network policy is active alongside the safety filter.
The plot in~\ref{fig:dagger_demo}(a) shows the trajectory over several laps from the first DAgger episode, where multiple instances of necessary safety filter intervention can be seen, as indicated by the color of the safety deviation norm.
In the early stages of training, the neural network policy has only been trained on the initial expert dataset, so it struggles to bring the vehicle onto the optimal racing line without trying to cut corners.
The safety filter must then deviate from applying the desired inputs to computing safe inputs that keep the vehicle in the track.
The plot in ~\ref{fig:dagger_demo}(b) shows the trajectory from the 5th episode, which is much more consistent than the initial policy with almost no major safety filter interventions.
The trajectory is aligned more closely with the optimal trajectory from MPCC, demonstrating an improved policy over previous iterations.

\addtolength{\textheight}{-1.5cm}   

\section{CONCLUSIONS}
In this work, we have presented a predictive safety filter that is able to render a closed-loop vehicle system safe when subject to any unsafe control signal.
A method for computing and verifying an invariant terminal set for the nonlinear vehicle system on constant curvature track segments is presented, providing a safe operating domain that does not overly restrict the desired policy.
The experiments illustrate two applications where the safety filter is able to ensure safety of the vehicle during dynamic high speed maneuvers.

\bibliographystyle{IEEEtran}
\bibliography{bibliography}

\end{document}

%% file: fig/tikz/controlLoopSafe.tex
\node [block] at (0,0) (rl) {Desired control signal $u_d(x(k))$};
\node [block] at (2.5,-2cm) (safecontroller) {Safety filter $\pi_s(x(k), u_d(k))$};
\node [block] at (-2.5,-2cm) (system) {System $f(x(k), u(k))$};
\draw [->,thick] (rl.east)  -- +(3cm,0) |- (safecontroller.east);
\draw [->,thick] (safecontroller.west)  |- (system.east);
\draw [->,thick] (system.west) -- +(-0.4cm,0) |- (rl.west);
\node at (4cm, -0.5cm) {$u_d(k)$};
\node at (-4cm, -0.5cm) {$x(k)$};
\node at (0cm, -1.6cm) {$u(k)$};

\draw[draw=black, dashed] (-5,-1) rectangle ++(10.1,-2.5);
\node at (2.2cm, -3.2cm) {Safe system $f_s(x(k), u_d(k))$};